\documentclass[
onecolumn,
]{ceurart}

\usepackage{graphicx}
\usepackage[caption=false]{subfig}
\usepackage{listings}
\usepackage{multirow}
\usepackage{makecell}
\mathchardef\mhyphen="2D
\newcommand\dlptrain{\mathit{D\mhyphen LP_{train}}}
\newcommand\dlptest{\mathit{D\mhyphen LP_{test}}}
\newcommand\dptrain{\mathit{D\mhyphen P_{train}}}
\newcommand\dptest{\mathit{D\mhyphen P_{test}}}

\begin{document}

\copyrightyear{2022}
\copyrightclause{Copyright for this paper by its authors.
  Use permitted under Creative Commons License Attribution 4.0
  International (CC BY 4.0).}

\conference{ISWC 2022: Deep Learning for Knowledge Graphs, October 23–27, 2022, Virtual Conference}

\title{Transformer-based Subject Entity Detection in Wikipedia Listings}

\author[1]{Nicolas Heist}[%
orcid=0000-0002-4354-9138,
email=nico@informatik.uni-mannheim.de,
url=http://www.uni-mannheim.de/dws/people/researchers/phd-students/nicolas-heist/
]
\cormark[1]
\author[1]{Heiko Paulheim}[%
orcid=0000-0002-4354-9138,
email=heiko@informatik.uni-mannheim.de,
url=http://www.heikopaulheim.com/
]

\address[1]{Data and Web Science Group, University of Mannheim, Germany}
\cortext[1]{Corresponding author.}

\begin{abstract}
In tasks like question answering or text summarisation, it is essential to have background knowledge about the relevant entities. The information about entities - and in particular, about long-tail or emerging entities - in publicly available knowledge graphs like DBpedia or CaLiGraph is far from complete. In this paper, we present an approach that exploits the semi-structured nature of listings (like enumerations and tables) to identify the main entities of the listing items (i.e., of entries and rows). These entities, which we call \textit{subject entities}, can be used to increase the coverage of knowledge graphs. Our approach uses a transformer network to identify subject entities on token-level and surpasses an existing approach in terms of performance while being bound by fewer limitations. Due to a flexible input format, it is applicable to any kind of listing and is, unlike prior work, not dependent on entity boundaries as input. We demonstrate our approach by applying it to the complete Wikipedia corpus and extract 40 million mentions of subject entities with an estimated precision of 71\% and recall of 77\%. The results are incorporated in the most recent version of CaLiGraph.
\end{abstract}

\maketitle

\section{Introduction}
\subsection{Motivation}
\label{motivation}
Background knowledge provides an essential advantage in tasks like text summarisation or question answering. With ready-to-use entity linking tools like Falcon \cite{sakor2020falcon}, entities in text can be identified and additional information can be drawn from background knowledge graphs (e.g. DBpedia \cite{lehmann2015dbpedia} or CaLiGraph\footnote{\url{http://caligraph.org}} \cite{heist2020entity}). Of course, this is only possible if the necessary information about the entity is included in the knowledge graph \cite{van2016evaluating}.

Hence, it is important to equip knowledge graphs with as much entity knowledge as possible. While this is easily possible for prominent entities that are mentioned frequently, the retrieval of information about long-tail and emerging entities that are mentioned only very infrequently is tedious \cite{farber2016emerging}. Still, approaches for automatic information extraction can be applied to increase the coverage of knowledge graphs to a certain extent. One strand of research is concerned with open information extraction systems that try to extract facts from web text (e.g. \cite{liu2020extracting,stanovsky2018supervised}). While they perform strongly on well-known entities, the extraction quality for long-tail entities is considerably worse \cite{liu2020extracting}.

\begin{figure}[t]
  \centering
  \includegraphics[width=.5\linewidth]{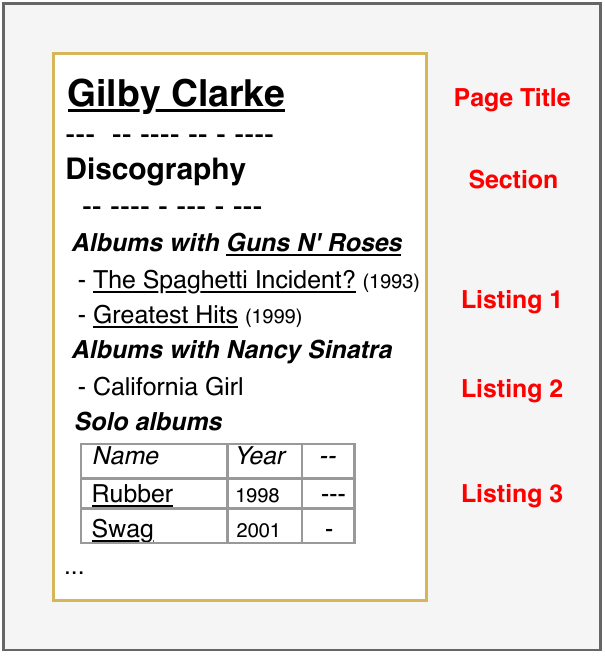}
  \caption{Simplified view on the listings of the Wikipedia page of Gilby Clarke.}
  \label{fig:running-example}
\end{figure}

The extraction of information from semi-structured data is in general less error-prone and has already proven to yield high-quality results as, for example, DBpedia itself is extracted primarily from Wikipedia infoboxes; other approaches use the category system of Wikipedia \cite{suchanek2007yago,heist2019uncovering,xu2016learning}; many more approaches focus on tables (in Wikipedia or the web) as a semi-structured data source to extract entities and relations (see \cite{zhang2020web} for a comprehensive survey).

In this work, we generalize over structures like enumerations (\textit{Listings 1 and 2}) and tables (\textit{Listing 3} in Figure~\ref{fig:running-example}) by simply considering them as listings with listing items (i.e., enumeration entries or table rows). Further, we call the main entity, that a listing item is about, a \textit{subject entity} (SE). In previous work, we defined SEs as \textit {all entities in a listing appearing as instances to a common concept} \cite{heist2021information}. In case of Figure~\ref{fig:running-example}, the SEs are the mentioned albums (e.g. \textit{The Spaghetti Incident?} or \textit{California Girl}). Here, the common concept is made explicit through the section labels above the listings (\textit{Albums with..}), but it may as well be the case that it is only implicitly defined through the respective SEs. As a listing item typically mentions only one SE together with some context (in this case, the publication year of the album), we assume that at most one SE per listing item exists.

In the English Wikipedia chapter alone, we find almost five million listings in roughly two million articles. From our estimation, about 80\% of the listings are suitable for the extraction of SEs, bearing an immense potential for knowledge graph completion (for details, see Section~\ref{listings-in-wikipedia}). Upon extraction, they can easily be digested by downstream applications: Due to the semi-structured nature of listings, the quality of extraction is higher than extraction from plain text, and SEs are typically extracted in groups of instances sharing a common concept (as given by the definition above). Especially the latter point makes subsequent disambiguation step much easier, as the group of extracted instances provides context for every individual instance. Another example of the downstream use of SEs is a work of ours where we used groups of SEs to learn lexical patterns that entail axioms \cite{heist2021information}. For example, if a listing is in a section that starts with \textit{Albums with}, we learn that the SEs are of the type \textit{Album}.

The combination of these two ideas, i.e. of extracting novel SEs and learning defining axioms for them, can bring a big benefit. In Figure~\ref{fig:running-example}, instead of simply discovering \textit{California Girl} as a new entity, we additionally assign the type \textit{Album}. Thinking further, we can learn an axiom that all albums mentioned in the discography of \textit{Gilby Clarke} are albums that are authored by him. The additional information can be used to refine the description of the extracted entity in the knowledge graph.

\subsection{Problem Statement}
\label{problem-statement}
Given an arbitrary listing, we want to identify the SEs among all entities mentioned in the listing. In the literature, there are only very few approaches that deal with this problem. The most related approach is a previous work of the authors that is concerned with the detection of SEs in Wikipedia list pages \cite{heist2020entity}.\footnote{List pages are special Wikipedia pages that contain only listings describing entities of a certain topic.} The approach uses a hand-crafted set of features to classify entities in tables or enumerations of list pages as SEs. However, the approach has several limitations:

\begin{itemize}
  \item It is only applicable to list pages and not to listings in any other context as the features are primarily designed for the list page context.
  \item Dependencies between individual SEs of listing items are not taken into account as the classification is done separately for every item.
  \item The approach needs mention boundaries of entities as input for the classification. Consequently, it cannot identify any new entities but only categorize existing entities into subject and non-subject entities.
\end{itemize}

\subsection{Contributions}
To harness the information expressed through SEs in more general settings, we aim to overcome the previously mentioned limitations in this work. In particular, we make the following contributions:

\begin{itemize}
    \item We present a Transformer-based approach for SE detection with a flexible input format that allows us to apply it to any kind of listing. Further, the model takes dependencies between listing items into account (Section~\ref{token-level-subject-entity-detection}).
    \item During prediction, the approach detects SEs end-to-end without relying on mention boundaries of the entities in the input sequence (Section~\ref{coarse-grained-entity-type-prediction}).
    \item We introduce a novel mechanism for generating negative samples of listings (Section~\ref{negative-sampling-through-shuffled-listing}) and a fine-tuning mechanism on noisy listing labels (Section~\ref{fine-tuning-on-noisy-page-labels}) leading to more accurate prediction results.
    \item In our evaluation, we show that the performance of our approach is superior to previous work (Section~\ref{evaluation-on-wikipedia-list-pages}); further, we analyse its performance in a more general scenario - that is, arbitrary listings of Wikipedia pages (Section~\ref{evaluation-on-wikipedia-page-listings}).
    \item We run the extraction of SEs on the complete Wikipedia corpus and incorporate the results in a new version of CaLiGraph (Section~\ref{subject-entity-extraction-over-wikipedia}).
\end{itemize}

The produced code is publicly available and part of the CaLiGraph extraction framework.\footnote{\url{https://github.com/nheist/CaLiGraph}}

\section{Related Work}
With the presented approach we detect SEs end-to-end, directly from listing text. For a given listing, we identify mentions of named entities and decide at the same time whether they are SEs of a listing or not. In the following, we first review Named Entity Recognition (NER) and subsequently discuss approaches that detect SEs.

\subsection{Named Entity Recognition}
NER is a subproblem of Entity Linking (EL) which only tries to identify mentions of named entities in the text without actually disambiguating them \cite{ling2015design}. As opposed to general Entity Recognition, NER only deals with the identification of named entities and ignores the linking of concepts (also called Wikification) \cite{milne2008learning}.

Early NER systems were based on hand-crafted rules and lexicons, followed by systems using feature-engineering and machine learning \cite{nadeau2007survey}. One of the first competitive NER systems that used neural networks has been presented by Collobert et al. in 2011 \cite{collobert2011natural}. This eventually lead to more sophisticated architectures based on word embeddings and LSTMs (e.g. from Lample et al. \cite{lample2016neural}).

With the rise of transformer networks \cite{vaswani2017attention} like BERT \cite{devlin2019bert} in 2018, they also found their direct application in NER (e.g. by Liang et al. \cite{liang2020bond}), or as part of an end-to-end EL system like the one from Broscheit \cite{broscheit2019investigating}. The latter uses a simple but effective prediction scheme, where entities are predicted at token-level and multiple subsequent tokens with the same predicted entity are collapsed into the actual entity prediction. In our work, we use a similar token-level prediction scheme to detect SEs.

\subsection{Subject Entity Detection}
Although SE detection has not explicitly been addressed in the literature very frequently, there are some approaches that deal with related problems or subproblems of it. In table interpretation, an important task is the identification of the \textit{subject column}, i.e. the column containing the entity with outgoing relations to all other columns. TAIPAN \cite{ermilov2016t} is an approach that aims to recover the semantics of tables and names subject column identification as the first major task towards relation extraction in tables. To identify subject columns, they choose the columns having entities with the most outgoing edges to entities in other columns w.r.t. a background knowledge graph. While this is a viable approach for tables that are already annotated with entities, it is not broadly applicable to general listings that may not have many known (or even annotated) entities.

Another related approach is from Zhao et al. \cite{zhao2021bert} who deal with a problem which they call \textit{key entity detection}. Primarily, they do sentiment analysis in financial texts and use the detection of key entities - which they define to be subjects of events related to financial information - in order to attribute the positive or negative sentiment to a concrete entity. Similar to our proposed approach, they use a Transformer to detect key entities. However, they only use it to select the key entities from a predefined set of entities and ignore the NER part.

As mentioned in the introduction, the most closely related approach is the authors' prior work \cite{heist2020entity}: using manually defined features and a binary XGBoost classifier, entities on list pages are classified into either subject entities or non-subject entities. For the page \textit{List of Japanese speculative fiction writers},\footnote{\url{https://en.wikipedia.org/wiki/List_of_Japanese_speculative_fiction_writers}} for example, all entities in the enumerations that are \textit{Japanese speculative fiction writers} are classified as SEs.

More concretely, the approach uses page features (e.g. number of sections or tables on the page), positional features (e.g. indentation level of entry in the enumeration), and linguistic features (e.g. whether the column header is synonymous with the list page title). Overall, SEs are extracted with a precision of 90\% and a recall of 67\%. The classifier is trained and evaluated with a set of list pages that are annotated through distant supervision with DBpedia for background knowledge. This part is discussed in detail in Section~\ref{distantly-supervised-training-data-generation-for-list-pages} as the approach presented here relies on this training data generation strategy as well.

\section{Preliminaries}
\subsection{Listings in Wikipedia}
\label{listings-in-wikipedia}
Overall, the English Wikipedia has more than five million articles. Roughly two million of them contain at least one listing in the form of an enumeration or a table. All over these pages, we find 3.5 million enumerations and 1.4 million tables.\footnote{These numbers exclude very small listings with less than three items, which we do not consider.} The roughly 90K list pages in Wikipedia contain the most structured and easily exploitable form of listings. Here, listings are almost exclusively used to list a number of entities that have some common property (e.g. all Japanese speculative fiction writers).

\begin{figure*}[t]
\subfloat[Listing containing no explicit mention of the entities (Source: \url{https://en.wikipedia.org/wiki/Sunrisers_Hyderabad_in_2018})]{%
  \fbox{\includegraphics[clip,width=.95\textwidth]{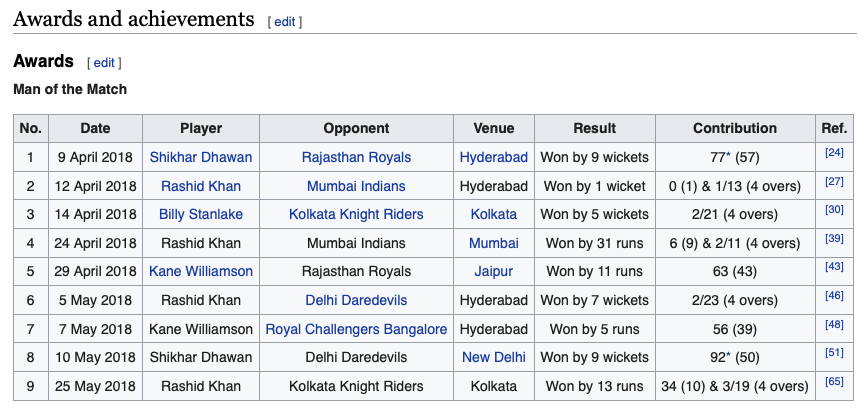}}%
  \label{fig:example-listing-1}
}
\hfill
\subfloat[Listing describing the properties of an entity (Source: \url{https://en.wikipedia.org/wiki/Dynamic_HTML})]{%
  \fbox{\includegraphics[clip,width=.95\textwidth]{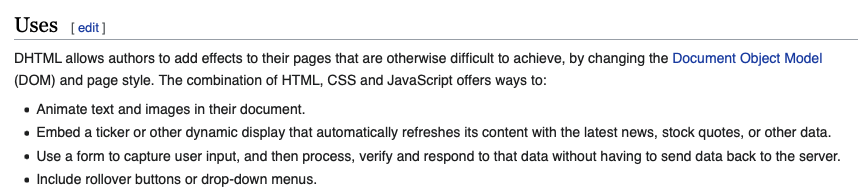}}%
  \label{fig:example-listing-2}
}
\hfill
\subfloat[Listing containing groups of entities (Source: \url{https://en.wikipedia.org/wiki/Ibiza_(Vino_de_la_Tierra)})]{%
  \fbox{\includegraphics[clip,width=.95\textwidth]{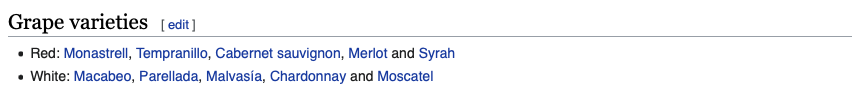}}%
  \label{fig:example-listing-3}
}
\caption{Examples of Wikipedia page listings with layout or content that is challenging for SE detection.}
\end{figure*}

Listings that appear on other Wikipedia pages are used for this purpose as well but not exclusively, which makes the detection of SEs much more complex. From the inspection of a sample of Wikipedia listings, we estimate that approximately 85\% of enumerations and 67\% of tables are usable for our approach. Especially enumerations are often used to simply structure content (e.g. to list the individual episodes in a biography). But even if listings are used to describe entities, they may not be usable due to various reasons:
\begin{itemize}
  \item Entity description without explicit mention (example in Figure~\ref{fig:example-listing-1})
  \item Description of the properties of a single entity  (example in Figure~\ref{fig:example-listing-2})
  \item Listing items contain groups of entities  (example in Figure~\ref{fig:example-listing-3})
\end{itemize}

Especially the first point renders a big portion of tables useless for our approach as an entity is implicitly described through entities and literals mentioned in multiple table columns (e.g. a sports match is described through date, player, opponent, and result).

\subsection{Distantly-Supervised Training Data Generation for List Pages}
\label{distantly-supervised-training-data-generation-for-list-pages}
In our experiments, we will use the training data generation strategy that we introduced in previous work \cite{heist2020entity}; to make this paper self-contained, we will give an overview of this strategy here. The strategy is based on the observation that DBpedia classes, Wikipedia categories, and Wikipedia list pages can be transformed into an immense taxonomy through linguistic and statistical methods. For example, the taxonomy contains the hierarchy \textit{Person > Writer > Speculative fiction writer > Japanese speculative fiction writer}. The first two elements originate from DBpedia classes, the third from a category, and the last from a list page.

As a consequence, we can use this hierarchy to infer the DBpedia classes of SEs for many list pages. To label the list page \textit{List of Japanese speculative fiction writers}, we assign every entity with the DBpedia class \textit{Writer} a positive label and every entity with a class that is disjoint with \textit{Writer} a negative label. Then we include all listing items into our training set that either have an entity with a positive label or only entities with negative labels. Other listing items are ignored as we cannot be certain that they may contain SEs which we could not identify due to the incompleteness of DBpedia.

The knowledge graph CaLiGraph \cite{heist2019uncovering,heist2020entity} uses this extended taxonomy of DBpedia classes, categories, and list pages as a type hierarchy, and enriches the original DBpedia instances with additional, more fine-grained types. Furthermore, CaLiGraph contains a higher number of instances than DBpedia as it additionally contains the extracted SEs from list pages.

\subsection{Transformers for Token Classification}
Pre-trained transformer networks \cite{vaswani2017attention} like BERT \cite{devlin2019bert} or DistilBERT \cite{sanh2019distilbert} produced new state-of-the-art results for various NLP tasks including NER and question answering. To a large extent, their ubiquitous application is due to the fact that only a comparably small amount of fine-tuning is necessary to fit them to various tasks. BERT, for instance, consists of 12 multi-head attention layers followed by a simple linear layer as classification head. To apply a transformer model to a token classification problem, it is oftentimes sufficient to fine-tune the final classification head. 

The input for a transformer model can consist of plain text and needs to be tokenized before it can be processed. Every word in the input sequence is transformed into one or more tokens (if the word is not contained in the vocabulary, multiple word-piece tokens are used). Further, the input sequence has to contain special tokens that indicate, for example, the start and the end of the sequence. Using BERT for token classification, the input sequence has a fixed length of 512 tokens, has to start with a \emph{[CLS]} token and end with a \emph{[SEP]} token. Additional special tokens may be introduced to provide more context information to the model.

\section{Subject Entity Detection with Transformers}
To detect SEs in listings, we phrase the problem as a token classification problem where we, similar to the work of Broscheit \cite{broscheit2019investigating}, produce a label for every token of the input sequence. In a subsequent step, we aggregate the token labels to predictions of SE mentions. We use 13 different token labels, such as \textit{Person} or \textit{Organisation}, to identify SEs and additionally make a prediction of their types (refer to Table~\ref{tab:subject-entity-extraction-results} for the full list of labels). In Section~\ref{token-level-subject-entity-detection} we explain how to create input sequences that preserve the context and the structure of a listing. In Section~\ref{coarse-grained-entity-type-prediction} we show our choice of labels for SE prediction, and in Section~\ref{negative-sampling-through-shuffled-listing} we introduce a mechanism to generate negative samples of listings. Finally, Section~\ref{fine-tuning-on-noisy-page-labels} explains how to use noisy SE labels on page listings for further fine-tuning of our models.

\subsection{Token-level Subject Entity Detection}
\label{token-level-subject-entity-detection}
To pass a listing for SE detection to the transformer model, we use multiple special tokens in order to encode context information (page, section, potential table header) and structural information (entries, rows, columns) of the listing into the input sequence. Every sequence consists of the listing context, followed by the special token indicating the end of context \textit{[CXE]}, and one or more listing items:

\begin{verbatim}
[CLS] <context> [CXE] <listing items> [SEP]
\end{verbatim}

We use the special token \textit{[CXS]} to separate context elements. Within listing items, table rows and columns are indicated with \textit{[ROW]} and \textit{[COL]}, respectively. For enumerations, we use the tokens \textit{[E1]} to \textit{[En]} to indicate the start of an entry with the indentation level 1 to n.

Ignoring that some words may be split into multiple tokens, the input for the first listing item of \textit{Listing 1} in Figure~\ref{fig:running-example} looks as follows:

\begin{verbatim}
[CLS] Gilby Clarke [CXS] Discography [CXS] Albums with Guns N' Roses [CXE]
[E1] The Spaghetti Incident? (1993) [SEP]
\end{verbatim}

We want the model to take dependencies between listing entities into account. For example, if the SE in the first listing item is mentioned right in the beginning, it is very likely that this is the case for the remaining listing items as well. Instead of only providing one listing item per input sequence, we can provide as many as the input sequence length permits. Through the attention layers within the Transformer architecture, the model is able to take these dependencies within the input sequence into account. Hence, we put \textit{Listing 1} into one input sequence:

\begin{verbatim}
[CLS] Gilby Clarke [CXS] Discography [CXS] Albums with Guns N' Roses [CXE]
[E1] The Spaghetti Incident? (1993)
[E1] Greatest Hits (1999) [SEP]
\end{verbatim}

Likewise, we encode \textit{Listing 3} as one input sequence:

\begin{verbatim}
[CLS] Gilby Clarke [CXS] Discography [CXS] Solo albums [CXS]
[ROW] Name [COL] Year [CXE]
[ROW] Rubber [COL] 1998
[ROW] Swag [COL] 2001 [SEP]
\end{verbatim}

If the listing is too long to fit into one input sequence, we split the listing items into chunks and process them one after another. Each chunk is augmented with the same context information and a different set of listing items. Depending on the length of listing items, it is possible to fit 20 or more items into one input sequence. In our ablation study in Section~\ref{ablation-study} we show that this item chunking strategy has a strongly positive effect on the recall of the model. But apart from that we immensely reduce the run time of the model for training and prediction. The number of processed input sequences is reduced by a factor that is roughly equivalent to the median number of items per listing.\footnote{We deliberately use the median and not the average of items per listing as large listings will be split into multiple input sequences due to the size limitation.}

\subsection{Coarse-grained Entity Type Prediction}
\label{coarse-grained-entity-type-prediction}
The most common notation to tag tokens in NER is the BIO notation (\textit{Begin}, \textit{Inside}, and \textit{Outside} of an entity) together with an entity type (e.g. \textit{Person} or \textit{Organisation}). We decided not to use the BIO notation as, per definition, there is at most one SE per listing item. Instead of making the task even simpler and getting rid of the entity type prediction in favor of a simple binary SE prediction task as well, we decided to stick with the coarse-grained entity type prediction. This has the advantage that the entity types can be used as additional information in downstream tasks - most importantly in a subsequent entity disambiguation step. In addition to that, we show in our ablation study in Section~\ref{ablation-study} that the more difficult task of entity type prediction even slightly increases the precision of the model. 

Context and special tokens are annotated with the \textit{IGNORE} label to indicate the model that we need no prediction for these tokens. SEs are annotated with the respective entity type, everything else is annotated with \textit{NONE}. Again ignoring word-piece tokenization, the labels for \textit{Listing 1} of Figure~\ref{fig:running-example} look as follows:

\begin{verbatim}
    IGNORE IGNORE IGNORE IGNORE IGNORE IGNORE IGNORE IGNORE IGNORE IGNORE IGNORE IGNORE
    IGNORE WORK_OF_ART WORK_OF_ART WORK_OF_ART NONE
    IGNORE WORK_OF_ART WORK_OF_ART NONE IGNORE
\end{verbatim}

\subsection{Negative Sampling through Shuffled Listings}
\label{negative-sampling-through-shuffled-listing}
It is difficult to find negative examples of complete listings if the training data is generated heuristically and with distant supervision as described in Section~\ref{distantly-supervised-training-data-generation-for-list-pages}. Positives can be found easily (i.e., there is an entity in the listing item that has the correct type), but the inverse does not always hold. If we do not find a positive, this may mean that the listing item does not contain one, but it is as well possible that the annotation is missing. From a logical standpoint, it is even unlikely that some items in a listing contain SEs while others do not.

To mitigate this problem, we equipped our approach with a sampling mechanism for negatives that randomly assembles them from the contexts and items of all positives in the training set. If the context and items are assembled randomly, the differences between the individual items (and the difference in the context) should be higher than in a real listing. The intention of this mechanism is that the model learns to identify the coherence between SEs of listing items as well as between items and the context.

For enumeration listings, the mechanism is simple as we pick the context from one listing and a random number of items (between three and the maximum number of items per chunk) from other listings. For table listings, we have to take care that the number of columns of an assembled listing is consistent. Hence, the positives from the training set are divided into groups of the same column size and listings are only assembled from within a single group. A negative example produced from four different listings could look as follows:

\begin{verbatim}
[CLS] Gilby Clarke [CXS] Discography [CXS] Albums with Guns N' Roses [CXE]
[E1] James Stewart as Billy Jim Hawkins
[E1] Curzon Mill Company, part of Ashton syndicate.
[E1] Brepholoxa Van Duzee, 1904 [SEP]
\end{verbatim}

The mechanism has exactly one hyper-parameter which is the proportion of negative listings to generate. We experiment with values between 0.0 (no negative samples at all) and 1.0 (as many negatives as we have positives).

\subsection{Fine-Tuning on Noisy Page Labels}
\label{fine-tuning-on-noisy-page-labels}
The training data generation strategy described in Section~\ref{distantly-supervised-training-data-generation-for-list-pages} lets us create labels for listings of list pages that we use for the initial training of our models. To train a model that works well on listings of any pages, additional training data of listings that are not on list pages may be beneficial (the differences in listings have been described in Section~\ref{listings-in-wikipedia}).

We gather this data by first training a model using the heuristically labelled list pages. We apply the model to listings of all pages for noisy labels of SEs. We then filter them by discarding any listings where multiple types of SEs have been predicted (e.g., if the first SE of a listing is labelled as \textit{PERSON} and the second is labelled as \textit{WORK\_OF\_ART}).

\section{Experiments}
The goal of our experiments is to compare the performance of our approach against previous work on SE detection in list pages (Section~\ref{evaluation-on-wikipedia-list-pages}) and evaluate its performance in the more general setting of Wikipedia page listings (Section~\ref{evaluation-on-wikipedia-page-listings}). Further, we analyze some of our design choices in an ablation study (Section~\ref{ablation-study}). Finally, we apply our best model to the complete Wikipedia corpus and report our extraction results (Section~\ref{subject-entity-extraction-over-wikipedia}).

\subsection{Metrics}
For the evaluation of our SE detection models, we stick to the common metrics for NER introduced in SemEval-2013 \cite{segura2013semeval}. We report precision, recall, and F1-scores of the following scenarios:
\begin{itemize}
    \item \textbf{Partial:} Prediction matches the boundary of the true entity at least partially.
    \item \textbf{Exact:} Prediction exactly matches the boundary of the true entity.
    \item \textbf{Ent-Type:} At least partial boundary match and entity type matches.
    \item \textbf{Strict:} Predicted boundary and type exactly match with the true entity.
\end{itemize}

\subsection{Datasets}
\begin{table*}
  \caption{Statistics of the datasets used for the experiments. The complete corpus $D$ contains all Wikipedia pages that have listings. $\dlptrain$ and $\dlptest$ are extracted from all Wikipedia list pages and are labelled through distant supervision; $\dptrain$ contains listings from arbitrary pages and contains noisy labels from a model trained on list pages while $\dptest$ is annotated manually.}
  \label{tab:dataset-statistics}
  \begin{tabular}{lccccccc}
    \toprule
    \textbf{Dataset} & \textbf{\#Pages} & \multicolumn{2}{c}{\textbf{\#Listings}} & \multicolumn{2}{c}{\textbf{Items per Listing (Avg.)}} & \multicolumn{2}{c}{\textbf{Items per Listing (Med.)}}\\
    & & \textbf{Enums} & \textbf{Tables} & \textbf{Enums} & \textbf{Tables} & \textbf{Enums} & \textbf{Tables}\\
    \midrule
    $D$ & 1,980,021 & 3,463,053 & 1,352,848 & 10.57 & 14.43 & \hphantom{1}6 & \hphantom{1}8\\
    \midrule
    $\dlptrain$ & \hphantom{1,0}68,494 & \hphantom{1,}289,666 & \hphantom{1,}116,715 & 18.06 & 31.26 & \hphantom{1}8 & 12\\
    $\dlptest$ & \hphantom{1,0}17,123 & \hphantom{1,0}75,063 & \hphantom{1,0}28,688 & 18.17 & 31.32 & \hphantom{1}8 & 12\\
    $\dptrain$ & \hphantom{1,}546,667 & \hphantom{1,}663,455 & \hphantom{1,}306,399 & 18.72 & 24.53 & 12 & 13\\
    $\dptest$ & \hphantom{1,000,}502 & \hphantom{1,000,}763 & \hphantom{1,000,}265 & \hphantom{1}8.42 & 11.25 & \hphantom{1}6 & \hphantom{1}7\\
  \bottomrule
\end{tabular}
\end{table*}

In the experiments, we primarily focus on Wikipedia as a data corpus due to its encyclopedic structure and the convenient mapping of entities to DBpedia and CaLiGraph. From the main dataset $D$ which consists of all Wikipedia pages that contain listings, we create the subsets $\dlptrain$ and $\dlptest$ (from list pages) as well as $\dptrain$ and $\dptest$ (from any pages with listings). The statistics of the datasets are shown in Table~\ref{tab:dataset-statistics}. For the experiments, we use a dump of the English Wikipedia from October 2020 to be compatible with the latest release of CaLiGraph.

The datasets $\dlptrain$ and $\dlptest$ are created as explained in Section~\ref{distantly-supervised-training-data-generation-for-list-pages}. For the experiments, we use a part of $\dlptrain$ for validation so that we have a distribution of 60\% training, 20\% validation, and 20\% test set (similar to \cite{heist2020entity}).

The datasets $\dptrain$ and $\dptest$ consist of listings from arbitrary Wikipedia pages. Hence, no type information is available to infer the SE labels through distant supervision. For $\dptrain$, we retrieved the labels as described in Section~\ref{fine-tuning-on-noisy-page-labels}. For $\dptest$, we provided the type information by manually annotating the roughly 1K listings with coarse-grained entity types (e.g. \textit{Person} or \textit{Organisation}). We mapped these types to their DBpedia counterparts and used this information to infer the SE labels via distant supervision. This substantially reduced the annotation effort from labelling roughly 10K listing items with concrete SE labels to labelling 1K listings with coarse-grained types. This implies that this dataset is also, in part, heuristically created and the results have to be taken with a grain of salt.

\subsection{Evaluation on Wikipedia List Pages}
\label{evaluation-on-wikipedia-list-pages}
\begin{table}
  \caption{Evaluation results for SE detection on Wikipedia list pages (evaluating on $\dlptest$). Precision, recall and F1-score (in \%) are given for the \textit{Exact} scenario. $Ours_{LP}$ is the best configuration for $\dlptest$ while $Ours_{P}$ is the best configuration for $\dptest$ using $\dlptrain$ as training data.}
  \label{tab:evaluation-on-wikipedia-list-pages}
  \begin{tabular}{lccccccccc}
    \toprule
    \textbf{Approach} & \multicolumn{3}{c}{\textbf{Enums}} & \multicolumn{3}{c}{\textbf{Tables}} & \multicolumn{3}{c}{\textbf{Overall}}\\
    & \textbf{P} & \textbf{R} & \textbf{F1} & \textbf{P} & \textbf{R} & \textbf{F1} & \textbf{P} & \textbf{R} & \textbf{F1}\\
    \midrule
    Heist and Paulheim \cite{heist2020entity} & 91 & 82 & 86 & \textbf{90} & 55 & 68 & 90 & 67 & 77\\
    $Ours_{LP}$ & \textbf{93} & \textbf{94} & \textbf{94} & 89 & \textbf{87} & \textbf{88} & \textbf{92} & \textbf{91} & \textbf{92}\\
    $Ours_{P}$ & 92 & 93 & 93 & 88 & 86 & 87 & 91 & 90 & 91\\
  \bottomrule
\end{tabular}
\end{table}

The evaluation results for experiments on the dataset $\dlptest$ are given in Table~\ref{tab:evaluation-on-wikipedia-list-pages}. We compare the approach Heist and Paulheim \cite{heist2020entity} with our model in the two configurations $Ours_{LP}$\footnote{Configuration: Model roberta-base trained for 3 epochs with batch size 64, learning rate 5e-5, no warmup or weight decay, negative sample size 0.5} and $Ours_{P}$.\footnote{Configuration: Model roberta-base trained for 2 epochs with batch size 64, learning rate 5e-5, no warmup or weight decay, negative sample size 0.3} Both configurations are trained with training part of $\dlptrain$ and tuned using the evaluation part. The former model configuration is the best one w.r.t. performance on $\dlptest$. The latter model configuration is the best one w.r.t. performance on $\dptest$. To train our models, we use the Huggingface transformer library \cite{wolf2020transformers}.

Both of our model configurations significantly outperform the existing approach Heist and Paulheim \cite{heist2020entity}, especially in terms of recall for both enumerations and tables, showing that our model can identify substantially more entities while keeping a high level of precision. For enumerations, the precision increased slightly and the recall is over ten percent higher. While precision is kept almost constant for tables, the recall increased by more than 30 percent.

\subsection{Evaluation on Wikipedia Page Listings}
\label{evaluation-on-wikipedia-page-listings}
\begin{table}
  \caption{Precision, recall and F1-score (in \%) for SE detection on Wikipedia page listings (evaluating on $\dptest$) using our best model configuration $Ours_{final}$.}
  \label{tab:evaluation-on-wikipedia-page-listings}
  \begin{tabular}{lccccccccc}
    \toprule
    \textbf{Metric} & \multicolumn{3}{c}{\textbf{Enums}} & \multicolumn{3}{c}{\textbf{Tables}} & \multicolumn{3}{c}{\textbf{Overall}}\\
    & \textbf{P} & \textbf{R} & \textbf{F1} & \textbf{P} & \textbf{R} & \textbf{F1} & \textbf{P} & \textbf{R} & \textbf{F1}\\
    \midrule
    Partial & 76 & 78 & 77 & 68 & 82 & 74 & 73 & 79 & 76\\
    Exact & 73 & 76 & 75 & 67 & 81 & 73 & 71 & 77 & 74\\
    Ent-Type & 76 & 78 & 77 & 65 & 78 & 71 & 73 & 78 & 75\\
    Strict & 71 & 74 & 73 & 64 & 77 & 70 & 69 & 75 & 72\\
    \midrule
    Baseline & 23 & 85 & 36 & 21 & 90 & 34 & 23 & 86 & 36\\
  \bottomrule
\end{tabular}
\end{table}

The evaluation results for the model $Ours_{final}$\footnote{Configuration: Similar to $Ours_{P}$ with an additional fine-tuning step of one epoch on D-$P_{train}$.} on $\dptest$ is given in Table~\ref{tab:evaluation-on-wikipedia-page-listings}. Comparing the \textit{Exact} scenario with the results on Wikipedia list pages, it becomes clear that the performance on arbitrary listings is worse. The losses in performance for tables are slightly higher than those for enumerations. This aligns with the observation that a lower portion of tables is usable for our approach. For tables, we have the advantage that mention boundaries are often indicated through column separators but this does not reflect in the results. In general, we notice that training the models for more than two to three epochs on $\dlptrain$ leads to overfitting on list page data and hence reduced performance on $\dptest$.

Unfortunately, it is not possible to apply the approach Heist and Paulheim \cite{heist2020entity} to this dataset as it contains several features that are specific to list pages. As an alternative, we implemented the pick-first-entity baseline which has already proven as a strong baseline in prior work \cite{heist2020entity}. In this baseline, we simply label the first mentioned entity in an item as SE. In Table~\ref{tab:evaluation-on-wikipedia-page-listings} we see that this baseline has a very high recall (as most SEs are mentioned in the beginning) while the precision is far lower than the one of $Ours_{final}$. This shows that the model is able to sort out many false positives (tripling precision) by sacrificing only some correct SEs. In cases where coverage is not the only important criterion (as is usually the case in knowledge graph completion), our model should be preferred. The more important point, however, is that our model does not depend on mention boundaries as input (which might also account for some loss in performance).

\subsection{Ablation Study}
\label{ablation-study}
\begin{table}
  \caption{Evaluation results for SE detection on Wikipedia page listings (evaluated on $\dptest$) for variations of our best model configuration $Ours_{final}$. Precision, recall and F1-score (in \%) are given for the \textit{Exact} scenario.}
  \label{tab:ablation-study}
  \begin{tabular}{lccccccccc}
    \toprule
    \textbf{Approach} & \multicolumn{3}{c}{\textbf{Enums}} & \multicolumn{3}{c}{\textbf{Tables}} & \multicolumn{3}{c}{\textbf{Overall}}\\
    & \textbf{P} & \textbf{R} & \textbf{F1} & \textbf{P} & \textbf{R} & \textbf{F1} & \textbf{P} & \textbf{R} & \textbf{F1}\\
    \midrule
    $Ours_{final}$ & \textbf{73} & 76 & \textbf{75} & \textbf{67} & 81 & \textbf{73} & \textbf{71} & 77 & \textbf{74}\\
    .. without item chunks & 70 & 35 & 47 & 63 & 40 & 49 & 68 & 37 & 48\\
    .. without type prediction & 69 & \textbf{78} & 73 & 54 & \textbf{84} & 66 & 64 & \textbf{79} & 71\\
    .. without negative sampling & 71 & 74 & 73 & 66 & 81 & 73 & 70 & 76 & 73\\
    .. without fine-tuning on pages & 65 & 48 & 55 & 67 & 64 & 66 & 66 & 52 & 58\\
  \bottomrule
\end{tabular}
\end{table}

To verify some assumptions that we made during the design of the SE detection approach, we perform an ablation study using the page listings dataset $\dptest$. Firstly, we investigate how much chunking of items in input sequences influences the performance of the model. The results in Table~\ref{tab:ablation-study} show that it has a slightly positive effect on precision (3\% for enumerations, 4\% for tables) and a roughly doubling effect on recall. The results confirm our assumption that the model is able to improve its predictions by considering the dependencies between the listing items.

Further, we investigate whether the additional prediction of entity types has an influence on the performance (as opposed to a binary prediction of SEs). The results show that there is a positive effect on precision and a slightly negative effect on recall. As the F1 measure increases slightly and as the predicted types provide additional information for downstream tasks, we stick with type prediction instead of binary SE prediction.

Additionally, we see from Table~\ref{tab:ablation-study} that our negative sampling mechanism slightly increases the precision and recall of our final model. Consequently, the model seems to be able to learn whether there is some consistency between the listing items in the input sequence.

Finally, the fine-tuning on pages has a very strong effect on recall as it comes with an increase of 25\% and the precision of the model is also increased by 5\%. This result confirms that additional fine-tuning on noisy labels still yields a huge benefit.

\subsection{Subject Entity Extraction over Wikipedia}
\label{subject-entity-extraction-over-wikipedia}
Applying the model $Ours_{final}$ to the complete dataset of Wikipedia listings $D$ took 13 hours on a single NVIDIA RTX A6000 GPU with 48GB of RAM. We extracted a total of 40 million entity mentions from 2.7M enumerations and 1M tables on 1.7 million pages. Of the 40 million entity mentions, 19.5 million can be traced back to 3.8 million known entities (i.e., the predicted mention boundary overlapped with an existing link in Wikipedia, and hence, CaLiGraph), which means that each known entity has on average 5.1 mentions. If we use that same factor of 5.1 to estimate the number of entities for the remaining 20.5 million entity mentions, they describe 4 million additional unknown entities that could be added to the knowledge graph.

\begin{table}
  \caption{Number of extracted mentions of subject entities for the whole Wikipedia dataset of listings $D$ aggregated by entity type.}
  \label{tab:subject-entity-extraction-results}
  \begin{tabular}{lrlrlr}
    \toprule
    \textbf{Entity Type} & \textbf{\#Mentions} & \textbf{Entity Type} & \textbf{\#Mentions} & \textbf{Entity Type} & \textbf{\#Mentions}\\
    \midrule
    PERSON & 13,622,704 & GPE & 1,519,747 & NORP & 230,707\\
    OTHER & 9,398,003 & PRODUCT & 1,000,117 & LANGUAGE & 86,354\\
    WORK\_OF\_ART & 7,148,235 & SPECIES & 964,922 & LAW & 11,490\\
    ORG & 2,916,528 & FAC & 893,226 & &\\
    LOC & 1,531,452 & EVENT & 370,440 & \textbf{Total} & \textbf{39,693,925}\\
    \bottomrule
  \end{tabular}
\end{table}

In Table~\ref{tab:subject-entity-extraction-results} we display the number of extracted entity mentions aggregated by entity type. Unsurprisingly, the most frequently extracted entities are of the types \textit{Person, Work of Art, Organisation, and Location}. Apart from that, the mention type distribution roughly resembles the distribution of entities in DBpedia \cite{heist2020knowledge}.

\section{Conclusion}
In this work, we have presented a Transformer-based SE detection approach that overcomes several limitations of prior work to make it applicable to more general settings and, at the same time, improve extraction performance. An evaluation of listings of Wikipedia pages shows that the performance for such a more general setting is considerably worse than for the scenario of Wikipedia list pages. While the inferior results can partly be attributed to conceptual limitations of SE detection in arbitrary listings (c.f. Section~\ref{listings-in-wikipedia}), further improvement is necessary so that the results can be consumed by downstream applications without extensive post-filtering.

We are developing a post-filtering mechanism that takes the differences within an extracted group of SEs into account. For example, we can discard a group of extracted SEs if their predicted entity types show a high degree of diversion.

In the extraction framework of CaLiGraph, we will integrate a subsequent entity disambiguation step, which matches the identified SE mentions with existing entities or creates new entities in the knowledge graph. The main challenge will be to match SEs with existing entities and at the same time match SEs with one another (as the same entity may be occurring in multiple listings).

Complementary to the disambiguation step, we plan to further enhance CaLiGraph by using the defining axioms extracted from the listing context. The disambiguation step can be supported by the information extracted from the axioms, and similarly, the disambiguated entities can help to refine the axiom extraction.

\bibliography{references}

\end{document}